\documentclass[12pt]{article}
\usepackage{epsf}

\title{ Online Learning with Ensembles
      }

 \author{R. Urbanczik  \\
        Neural Computing Research Group\\
        Aston University\\ Aston Triangle, Birmingham B4 7ET, U.K.
       }

\newcommand{\R}{I\!\!R}
\newcommand{\sign}{ \mbox{\rm sign} }

\newcommand{\half}{{\frac{1}{2}}}

\begin{document}
\maketitle

\begin{abstract}
 \setlength{\parindent}{0.0cm}
 Supervised online learning with an ensemble of students randomized by 
 the choice of initial conditions is analyzed. For the case of the 
 perceptron learning rule,  asymptotically the same improvement in the
 generalization error of the ensemble compared to the performance of
 a single student is found as in Gibbs learning. For more optimized
 learning rules, however, using an ensemble yields no improvement.
 This is explained by showing that for any learning rule $f$ a transform
 $\tilde{f}$ exists, such that a single student using $\tilde{f}$ has
 the same generalization behaviour as an ensemble of $f$-students.
\end{abstract}

Online learning, where each training example is presented just once
to the student, has proved to be a very successful paradigm in the study
of neural networks using methods from statistical mechanics \cite{Saa98}.
On the one hand, it makes it possible to rigorously \cite{Ree98} analyze
a wide range of learning algorithms. On the other hand, online algorithms 
can in some cases yield a performance which equals that of the Bayes
optimal inference procedure,
e.g. asymptotically, when the probability of the data is a smooth function
of the parameters of the network \cite{Opp96}. 

Some problems, however, do remain. For nonsmooth cases, which arise e.g. 
in classification tasks, the Bayes optimal procedure yields a 
superior generalization performance, even asymptotically, to that
of online algorithms \cite{Opp91,Kin92}. Also even for smooth problems, 
the online
dynamics often has suboptimal stationary points arising from symmetries
in the network architecture. Then the sample size needed to reach the 
asymptotic regime will scale faster than linearly with the number of free 
parameters if no prior knowledge is built into the
initial conditions of the dynamics \cite{Bie96}. 

It thus seems of interest to ask which extensions of the online framework
make sense. Here we shall consider using an ensemble of students randomized
by the choice of initial condition and classifying a new input by a majority 
vote. This may be motivated by the fact that in the batch case the Bayes 
optimal inference procedure can be implemented by an ensemble 
picked from the posterior (given the training set) distribution on the
set of all students. We shall consider an ensemble of $K$ students;  
at time step $\mu$ the $i$-th student is characterized by a weight
vector $J_i^\mu\in \R^N$. The
learning dynamics is based on a training set of $\alpha N$ input/output pairs 
$(\xi^\mu,\tau^\mu)$ where $\xi^\mu\in \R^N$. 
We shall consider
realizable learning in a perceptron, so 
$\tau^\mu = \sign(B^T\xi^\mu)$ where $B$ is the $N$-dimensional weight 
vector defining the teacher and it is convenient to assume that $|B|=1$
holds for the Euclidean norm.
The dynamics of the $i$-th student then takes the form
\begin{equation}
J_i^{\mu+1} = 
  J_i^\mu +   \xi^\mu N^{-1} 
              f(\mu/N,|J_i^\mu|,B^T\!\xi^\mu,{J_i^\mu}^T\!\xi^\mu) 
\label{ensup}
\end{equation}
and the choice of the real valued function $f$ defines the learning rule.
Reasonably $f$ may only depend on the third argument $B^T\!\xi^\mu$ via its
sign $\tau^\mu$, but it is not helpful to make this explicit in the notation.
Note that all of the members of the ensemble learn from 
the same training examples, and these are presented in the same order.

Assuming that the components of the example inputs are independent random
variables picked from the normal distribution on $\R$, the state of the
ensemble can be described by the order parameters 
$R_i(\alpha) = B^T\! J_i^{\alpha N}$ and 
$Q_{ij}(\alpha) = {J_i^{\alpha N}}^T\!J_j^{\alpha N}$. For a reasonable
choice of $f$ \cite{Ree98} the order parameters will be nonfluctuating for 
large $N$ and satisfy the differential equations:
\begin{eqnarray}
\dot{R_i} &=& \langle y f_i^\alpha \rangle_{x_i,y} \nonumber \\
\dot{Q}_{ij} &=& \langle 
                    x_i f_j^\alpha + x_j f_i^\alpha + f_i^\alpha f_j^\alpha
                   \rangle_{x_i,x_j,y}  \nonumber \\
f_i^\alpha &\equiv& f(\alpha,Q^\half_{ii},y,x_i), \label{sys}
\end{eqnarray}
where $y$ and the $x_i$ are zero mean Gaussian random variables with
covariances $\langle x_i y \rangle = R_i$ and 
$\langle x_i x_j \rangle = Q_{ij}$. We shall only consider the case
where the initial values $J_i^0$ are picked independently from the uniform
distribution on a sphere with radius $\sqrt{P(0)}$. Then for large $N$
the initial conditions for (\ref{sys}) are $R_i(0)=Q_{ij}(0)=0$ for $i\neq j$
and $Q_{ii}(0) = P(0)$. These conditions are invariant under permutations of 
the site indices $i$ and this also holds for the system of differential 
equations (\ref{sys}). Thus this site symmetry will be preserved for all times
and we need only consider the three order parameters 
$R(\alpha) = R_i(\alpha)$, $P(\alpha) = Q_{ii}(\alpha)$ and 
$Q(\alpha) = Q_{ij}(\alpha)$ for $i\neq j$. Since the length of the students
is of little interest, it will often be convenient to consider the normalized
overlaps $r(\alpha) = R(\alpha)/\sqrt{P(\alpha)}$ and 
$q(\alpha) = Q(\alpha)/P(\alpha)$.

A new input $\xi$, picked from the same distribution as the training inputs,
will be classified by the ensemble using a majority vote, that is by:
\begin{equation}
\sigma(\xi) = \sign( \sum_{i=1}^K \sign({J_i^{\alpha N}}^T\!\xi)). 
\label{ensout}
\end{equation}
As an alternative to using a majority vote, one might consider constructing
a new classifier by averaging the weight vectors of the students, setting
$\bar{J}^{\alpha N} = K^{-1}\sum_i J_i^{\alpha N}$. 
As in Gibbs theory \cite{Wat93a} a simple application of the law of large
numbers yields that the two classifiers are equivalent in the large $K$ limit
if $q(\alpha)={\cal O}(1)$, that is 
$\sigma(\xi) = \sign(\bar{J}^{\alpha N} \xi)$ for almost all inputs.
In the sequel we shall only consider the large $K$ limit, assuming that 
$K\ll N$ so that the fluctuations in the site symmetry of the initial 
conditions 
can be ignored. The generalization error $\epsilon_{\rm e}$ of the ensemble,
that is the probability of misclassifying $\xi$, is then given by
$\epsilon_{\rm e} = \epsilon(r(\alpha)/\sqrt{q(\alpha)})$ where
\begin{equation}
\epsilon(x) = \frac{1}{\pi} \arccos x\;.
\end{equation}
Similarly, the generalization error of a single student in the ensemble is 
$\epsilon_{\rm s} = \epsilon(r(\alpha))$.

We shall first consider a soft version of the perceptron learning rule:
\begin{equation}
f = \eta |J_i^\mu| 
     H\!\left(\tau^\mu \frac{k}{\sqrt{1-k^2}}
                       \frac{{J_i^\mu}^T\xi^\mu}{|J_i^\mu|}\right) 
     \tau^\mu\;, \label{softp}
\end{equation}
where $H(x) = \frac{1}{2}{\rm erfc}(x/\sqrt{2})$ and $\eta$ is a 
time dependent learning rate. For $k=0$ this reduces to Hebbian learning 
whereas $k=1$ yields the perceptron learning rule. Note, however, that
the $|J_i^\mu|$ prefactor makes the dynamics invariant with respect to the
scaling of the student weight vectors. From (\ref{sys}) one obtains for
the order parameters:
\begin{eqnarray}
\dot{r} &=& \frac{\eta}{\sqrt{2 \pi}}(1-r^2) - 
             \frac{\eta^2}{2}r\left(\epsilon(kr) - 
                               \frac{1}{2}\epsilon( k^2)\right) 
\nonumber \\
\dot{q} &=& \frac{2 \eta}{\sqrt{2 \pi}}r(1-q) +
             \eta^2 \left(
               (1-q)\epsilon(kr) - 
               \frac{1}{2}\epsilon( k^2q) + 
               \frac{q}{2}\epsilon( k^2)    \right)\;. \label{soft}
\end{eqnarray}

We first consider the perceptron learning rule i.e. $k=1$. In the limit
$r,q \rightarrow 1$ one finds 
$\dot{r} \sim \eta \sqrt{2/\pi}(1-r) - \eta^2 \epsilon(r)/2$ and
$\dot{q} \sim \eta \sqrt{2/\pi}(1-q) - \eta^2 \epsilon(q)/2$, that is $r$ and
$q$ satisfy the same differential equation. If the learning rate schedule
is such that this limit is reached, this means that
that $(1-r)/(1-q)$ will approach $1$ for large 
$\alpha$. Hence asymptotically  
$\epsilon_{\rm e} \sim \epsilon(\sqrt{r(\alpha)})$, and the same improvement
by a factor of $1/\sqrt{2}$ in the generalization error of the ensemble
compared to single student performance is found as in Gibbs learning.
(Interestingly the same asymptotic relationship between $\epsilon_{\rm e}$ 
and $\epsilon_{\rm s}$ also holds for the Adatron learning rule 
$f =- \Theta(-\tau^\mu {J_i^\mu}^T\!\xi^\mu){J_i^\mu}^T\!\xi^\mu$).
The optimal asymptotics of the learning rate schedule is 
$\eta \sim 2\sqrt{2 \pi}/\alpha$ and this yields an 
$\epsilon_{\rm e} \sim \frac{2\sqrt{2}}{\pi\alpha} \approx 0.90/\alpha$
decay of the ensemble generalization error. This is very close to the
$0.88/\alpha$ decay found for the optimal single student algorithm
\cite{Kin92}.

We next consider improving the performance by tuning $k$. From (\ref{soft})
one easily sees that single student performance is optimized when 
$k = r$. Asymptotically this may be achieved by setting 
$k \sim 1-4/\alpha^2$ and choosing the optimal learning schedule which
is asymptotically the same as for the standard perceptron learning rule. 
Then already a single student achieves 
$\epsilon_{\rm s} \sim \frac{2\sqrt{2}}{\pi\alpha}$,
that is the same large $\alpha$ behaviour as the ensemble in the
$k=1$ case. Unfortunately  $r$ and $q$ now have a different asymptotics
and one finds $1-q \ll 1-r$. So for all practical purposes the ensemble 
collapses to a single point and for large $\alpha$ to leading order
$\epsilon_{\rm e} \sim \epsilon_{\rm s}$.

It is of course not clear that optimizing single student performance is
a good idea, and we thus analyze more generic schedules, setting 
$k \sim 1 - (\lambda/\alpha)^2$. Figure 1 then, however, shows that the 
two case considered above are optimal for ensemble and respectively single
student performance.

\newcommand{\tJ}{\tilde{J}}

The above analysis of the soft perceptron rule suggests that while for some
rules using an ensemble does significantly improve on single student 
performance, for
more optimized rules this may no longer be the case. We shall now prove that 
the generalization error of the optimal single student learning rule
is also a lower bound of the ensemble performance for any learning
rule $f$. To achieve this, a learning rule $\tilde{f}$  will be given which
for each pattern yields the ensemble average of $f$. Then a single student 
$\tJ^\mu$ using $\tilde{f}$ will have generalization behaviour 
equal to that of a large ensemble of students using $f$.
The dynamics for $\tJ^\mu$ may be written as
\begin{equation}
\tJ^{\mu+1} = \tJ^\mu + 
 \xi^\mu N^{-1} \tilde{f}(\mu/N,B^T\!\xi^\mu,
                          \mbox{$\tJ^{\mu}$}^T\!\xi^\mu)
\end{equation}
where $\tilde{f}$ is the following integral transform of $f$:
\begin{equation}
\tilde{f}(\alpha,y,\tilde{x}) = 
\left\langle 
  f(\alpha,P(\alpha)^\half,y,\tilde{x}+(P(\alpha)-Q(\alpha))^\half z)
\right\rangle_z\;. \label{tilde}
\end{equation}
Here the distribution of $z$ is normal. The entire procedure is quite 
intuitive: $\tJ^\mu$ represents the center of mass of the ensemble
and $\mbox{$\tJ^{\mu}$}^T\!\xi^\mu+(P(\alpha)-Q(\alpha))^\half z$
is a guess for the value of the hidden field ${J_i^\mu}^T\!\xi^\mu$ of
one of the ensemble members. For large $K$ the distribution of the last two 
quantities will be the same, and the ensemble average of $f$ can be reliably
predicted. Further, note that the class of soft perceptron rules (\ref{softp})
is invariant  under the integral transform (\ref{tilde}) since
$\langle H(a+bz) \rangle_z = H(a/\sqrt{1+b^2})$. This explains why 
optimizing single student and optimizing ensemble performance within
this class yields the same generalization behaviour.

To demonstrate that $\tJ^\mu$ does indeed emulate the large ensemble  
consider the order parameters $\tilde{R}(\alpha) = B^T\!\tJ^{\alpha N}$ 
and $\tilde{Q}(\alpha) =  |\tJ^{\alpha N}|^2$. We shall start with
$\tJ^0 = 0$ , thus $\tilde{R}(0) = R(0) = \tilde{Q}(0) = Q(0)=0$, and
it will suffice to show that the pair $\tilde{R},\tilde{Q}$
satisfies an identical differential equation as the pair $R,Q$. From 
(\ref{sys}) we obtain for Q:
\begin{equation}
\dot{Q} = \left\langle 2 x_i f(\alpha,P(\alpha)^\half,y,x_j) +
                  f(\alpha,P(\alpha)^\half,y,x_i)
                  f(\alpha,P(\alpha)^\half,y,x_j) \right\rangle_{y,x_i,x_j}
\label{dQ}
\end{equation}
where $i$ and $j$ are any two different indices. The Gaussians  $x_i$ and 
$x_j$ may be rewritten in terms of normal random variables $z_i,z_j$ and $z$,
independent of each other and of $y$, as
\begin{equation}
x_i = \sqrt{P-Q}z_i + \sqrt{Q-R^2}z + R y \mbox{ and }
x_j = \sqrt{P-Q}z_j + \sqrt{Q-R^2}z + R y \;.
\end{equation}
Carrying out the integrations over $z_i$ and $z_j$ in (\ref{dQ}) yields
\begin{equation}
\dot{Q} = \left\langle 2 \tilde{x} \tilde{f}(\alpha,y,\tilde{x}) +
                  \tilde{f}(\alpha,y,\tilde{x})^2
          \right\rangle_{y,z}\;,
\end{equation}
where $\tilde{x} \equiv \sqrt{Q-R^2}z + R y$. The variance of $\tilde{x}$
is $Q$ and its covariance with $y$ is $R$. Applying (\ref{sys}) to $\tJ^\mu$
yields 
$
\dot{\tilde{Q}} = \left\langle 2 \tilde{x} \tilde{f}(\alpha,y,\tilde{x}) +
                  \tilde{f}(\alpha,y,\tilde{x})^2
          \right\rangle_{y,\tilde{x}}
$, where the variance of $\tilde{x}$ is $\tilde{Q}$ and its covariance
with $y$ is $\tilde{R}$. Thus $Q$ and $\tilde{Q}$ satisfy the same differential
equation and an analogous argument shows that the same holds for $R$ and 
$\tilde{R}$.

It is interesting to ask whether the above equivalence between ensemble and 
single student behaviour carries over to more general situations. Let us 
first consider allowing interactions between the ensemble members. In this
case much more complicated scenarios can arise. However, if one only 
considers global and symmetric interactions between the ensemble members,
an equivalent single student rule will often exist. To be specific assume that
$f$ may in addition depend on the output of the entire ensemble 
(\ref{ensout}). This just amounts to allowing $f$ to depend on the 
random variable $\tilde{x}$, and with only minor 
modifications the above construction will again 
yield an equivalent single student rule.

Next consider more general architectures than the simple perceptron.
It is straightforward to generalize the construction to the case
of a tree committee machine: one just has to carry out an integration
analogous to (\ref{tilde}) per branch of the tree. The case of the tree
parity machine, however, is more involved  since due to a gauge symmetry,
students with differing weight vectors can implement the same function.
Thus averaging the output of the ensemble members (\ref{ensout}) 
may no longer be equivalent to averaging the weight vectors. But it is 
straightforward to break the symmetry in a formal way by adding a small 
deterministic drift term of the form $B\delta N^{-1}$ to the update 
equations (\ref{ensup})
of each branch. Then for $\delta>0$ the same procedure as for the 
tree committee will yield an equivalent single student rule. In the end,
one will of course want to take the limit $\delta\rightarrow 0$.
In this limit, however, for a training set size which is on the order
of the number of free parameters in a single student, 
only a trivial generalization behaviour
will result \cite{Sim96}. So this procedure does not allow us to make any 
statement about the  equivalence between ensemble and single student 
performance for the large training sets needed to achieve a nontrivial
behaviour. It does, however, show that the pathological divergence of the 
training times which results from the symmetry, cannot be overcome by the use 
of an ensemble. Similar remarks as for the tree parity machine apply to
fully connected architectures.

So in sharp contrast to batch learning where ensemble performance is often 
superior to single student performance, in online learning one cannot improve
on optimal single student performance through an ensemble. But obviously
if the state space of the learning system where large enough to store the 
entire training set, online learning would reduce to the batch case. So an
ensemble may simply not be an effective way making use of a state space
which is larger than in the case of a single student, and future research 
should investigate more efficient strategies of utilizing a large state space.

It is a pleasure to acknowledge helpful discussions with Manfred Opper and
David Saad.

\bibliographystyle{plain}
\bibliography{../tex/neural}

\begin{thebibliography}{1}

\bibitem{Bie96}
M.~Biehl, P.~Riegler, and C.~W{\"o}hler.
\newblock Transient dynamics of on-line learning in two-layered neural
  networks.
\newblock {\em J. Phys. A}, 29:4769, 1996.

\bibitem{Kin92}
O.~Kinouchi and N.~Caticha.
\newblock Optimal generalization in perceptrons.
\newblock {\em J. Phys. A}, 25:6243, 1992.

\bibitem{Opp96}
M.~Opper.
\newblock On-line versus off-line learning fron random examples: General
  results.
\newblock {\em Phys. Rev. Lett.}, 77:4671 --4674, 1996.

\bibitem{Opp91}
M.~Opper and D.~Haussler.
\newblock Generalization performance of {B}ayes optimal prediction algorithm
  for learning a perceptron.
\newblock {\em Phys. Rev. Lett.}, 66:2677--2680, 1991.

\bibitem{Ree98}
G.~Reents and R.~Urbanczik.
\newblock Selfaveraging and on-line learning.
\newblock {\em Phys. Rev. Lett.}, 80:5445 --5448, 1998.

\bibitem{Saa98}
D.~Saad, editor.
\newblock {\em On-line Learning in Neural Networks}.
\newblock Cambridge University Press, 1998.

\bibitem{Sim96}
R.~Simonetti and N.~Caticha.
\newblock On-line learning in parity machines.
\newblock {\em J. Phys. A}, 29:4859--4867, 1996.

\bibitem{Wat93a}
T.~Watkin.
\newblock Optimal learning with a neural network.
\newblock {\em Europhys. Lett.}, 21:871 -- 876, 1993.

\end{thebibliography}

\begin{figure}[p]
    \begin{center}\begin{tabular}[b]{c}
        \epsfbox{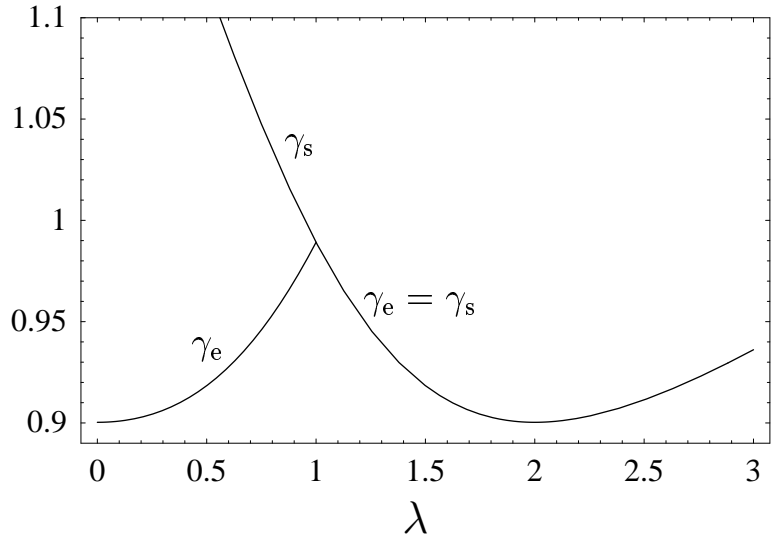}\\\
        \hspace*{0.5cm}\Large $\lambda$
   \end{tabular}\end{center}
 \caption{Asymptotics of the soft perceptron learning rule.
         The generalization error of the ensemble decays as 
         $\epsilon_{\rm e} \sim \gamma_{\rm e}/\alpha$, and for 
         a single student 
         $\epsilon_{\rm s} \sim \gamma_{\rm s}/\alpha$. The dependence of
         $\gamma_{\rm e}$ and $\gamma_{\rm s}$ on the parameter $\lambda$
         which controls the softness of the learning rule via 
         $k \sim 1 - (\lambda/\alpha)^2$, is shown in the plot. The  
         learning rate schedule is $\eta \sim 2\sqrt{2 \pi}/\alpha$.
         For all values of $\lambda$, this schedule optimizes both single
         student and ensemble performance.
         For $\lambda > 1$ the students in the ensemble correlate quickly
         with increasing $\alpha$, and using an ensemble
         asymptotically yields no improvement over single student performance.
         }
\end{figure}

\end{document}